\documentclass[9pt,conference]{IEEEtran}
\IEEEoverridecommandlockouts

\usepackage{cite}
\usepackage{amsmath,amssymb,amsfonts}
\usepackage{algorithmic}
\usepackage{graphicx}
\usepackage{textcomp}
\usepackage{xcolor}
\usepackage{eso-pic}
\usepackage{booktabs}
\usepackage{subcaption}
\usepackage{color}
\usepackage{multirow}
\usepackage{hyperref}
\usepackage{pifont}
\newcommand{\cmark}{\ding{51}}%

\renewcommand{\IEEEtitletopspaceextra}{10mm} 


\AddToShipoutPictureBG*{%
  \ifnum\value{page}=1  
    \put(10,780){\color{black}\scriptsize \copyright\ 2024 IEEE. Personal use of this material is permitted. Permission from IEEE must be obtained for all other uses, in any current or future media, including reprinting, republishing this material for}
    \put(10,770){\color{black}\scriptsize advertising or promotional purposes, creating new collective works, for resale or redistribution to servers or lists, or reuse of any copyrighted component of this work in other works.}
  \fi
}

\def\BibTeX{{\rm B\kern-.05em{\sc i\kern-.025em b}\kern-.08em
    T\kern-.1667em\lower.7ex\hbox{E}\kern-.125emX}}
\begin{document}

\title{SoundBeam meets M2D: Target Sound Extraction with Audio Foundation Model
}

\author{
    \IEEEauthorblockN{Carlos Hernandez-Olivan, Marc Delcroix, Tsubasa Ochiai, Daisuke Niizumi, \\Naohiro Tawara, Tomohiro Nakatani, Shoko Araki
}
    \IEEEauthorblockA{
        NTT Corporation, Japan
    }
    \thanks{This work was done during an internship at NTT. The work was supported by JST Strategic International Collaborative Research Program (SICORP), Grant Number JPMJSC2306, Japan.}
    }
    

\maketitle

\begin{abstract}
Target sound extraction (TSE) consists of isolating a desired sound from a mixture of arbitrary sounds using clues to identify it. A TSE system requires solving two problems at once, identifying the target source and extracting the target signal from the mixture.  For increased practicability, the same system should work with various types of sound. The duality of the problem and the wide variety of sounds make it challenging to train a powerful TSE system from scratch. In this paper, to tackle this problem, we explore using a pre-trained audio foundation model that can provide rich feature representations of sounds within a TSE system. We chose the masked-modeling duo (M2D) foundation model, which appears especially suited for the TSE task, as it is trained using a dual objective consisting of sound-label predictions and improved masked prediction. These objectives are related to sound identification and the signal extraction problems of TSE. We propose a new TSE system that integrates the feature representation from M2D into SoundBeam, which is a strong TSE system that can exploit both target sound class labels and pre-recorded enrollments (or audio queries) as clues. We show experimentally that using M2D can increase extraction performance, especially when employing enrollment clues.
\end{abstract}


\section{Introduction}
\label{sec:intro}
\renewcommand{\thefootnote}

Target sound extraction (TSE) is a technology that isolates a desired source signal in a mixture of arbitrary sounds. This technology can have many applications for sound production, hearables, etc. \cite{semhear}. TSE can be realized with a neural network that takes the sound mixture as input and outputs the target sound signal, conditioned on target sound embeddings that identify the target in the mixture. We can derive the target sound embeddings from class labels \cite{ochiai20_interspeech,soundbeam}, audio enrollments (audio queries) of the target sound \cite{soundbeam, gfeller2021one}, or both \cite{soundbeam}. Other works have also exploited other clues, such as video \cite{ZhaoGRVMT18} or language clues \cite{clapsep,OkamotoHYIK22}, but these are beyond our scope.

The Class label clues are represented by a one-hot vector, which is mapped to an embedding space
representing the different sound classes using an embedding layer. Such class label-based TSE systems can directly learn the target sound embeddings optimal for TSE. However, such TSE systems cannot extract sounds unseen during training. In contrast, enrollment clue-based TSE maps the audio enrollment to the embedding space using a clue encoder network, thus extracting sounds that share similar characteristics to the enrollment sample without explicitly relying on SE class labels. They can thus generalize to sounds unseen during training. Recently, we proposed a system, SoundBeam \cite{soundbeam}, which is a TSE system 
trained by alternating between class label and enrollment clues, combining thus the advantages of both clues as well as improving extraction performance thanks to the multi-task training effect.

Training a TSE system is challenging because it needs to learn a representation of sounds that enables it to identify them in a mixture while also learning to extract the target sound signal. This is especially hard when learning the problem from scratch, as it requires training the TSE with a very large amount of sound combinations to allow the model to generalize to the extraction of many sound classes.

Recently, foundation models, such as self-supervised learning (SSL)-based models have received increased attention to facilitate the training of complex tasks. Foundation models learn rich general-purpose feature representations from vast amounts of data, which can then be used to learn task-specific models. For example, pre-trained SSL models, or models trained on vast amounts of data, such as Whisper\cite{whisper}, have been widely used for automatic speech recognition (ASR) \cite{mohamed2022self}, diarization \cite{wavlm}, speech enhancement \cite{junyi}, etc. Beyond speech, foundation models have also been explored to obtain general representations of audio signals and shown promising results for, e.g., sound event classification \cite{pann,ast,byol-a}. In particular, the recently proposed masked-modeling duo (M2D) model \cite{m2d_taslp}, which is trained on the AudioSet \cite{audioset} using a combination of a \emph{modified masked auto-encoder} (MAE)\cite{mae} and \emph{sound label classification} training objectives, has been shown to achieve state-of-the-art performance for various audio processing tasks \cite{m2d_taslp}. 

In this paper, we explore the potential of using the strong audio foundation model, M2D, for TSE. We hypothesize that the M2D model trained with its dual objective is particularly well-suited for the TSE task. Indeed, TSE must 1) identify the target sound in the mixture, which is related to the M2D model sound label classification training objective, and 2) output the target sound signal, which is related to the MAE training objective. 
We propose a new TSE system that combines SoundBeam and M2D. We employ the pre-trained M2D model to 1) generate the target sound embedding from the enrolment clue, and 2) generate a representation of the mixture that serves as additional input for SoundBeam. For the latter, we introduce an adaptive input enhancer (AIE) module, to combine the feature representation of the TSE encoder and M2D model. 

We perform TSE experiments using noisy and reverberant sound mixtures. We reveal that exploiting the representation from the M2D model for both the enrollment and the mixture greatly improves performance, especially when using enrollment clues because it allows obtaining more discriminative target sound class embeddings. These results demonstrate the potential of using general audio foundation models such as M2D for TSE. 

The main experiments are performed with SoundBeam, which is an offline TSE system. 
However, many applications require online processing. Therefore, we also perform some experiments confirming that M2D can also benefit causal and lightweight TSE systems, such as Waveformer \cite{waveformer}. 

\section{Related Works}
Pre-trained foundation models have been used in downstream tasks such as speech enhancement \cite{HuangWYGK22}, target speech extraction \cite{junyi} or TSE \cite{ clapsep,zhao2024universalsoundseparationselfsupervised}. 
Concurrent with our work, recent pre-prints \cite{clapsep, zhao2024universalsoundseparationselfsupervised } have also explored exploiting audio foundation models for TSE. In \cite{ clapsep}, the contrastive language-audio pre-trained (CLAP) model \cite{LAION-CLAP} was used to obtain target sound embeddings from text descriptions (audio captions), enrollment clues, and to process the input mixture. In \cite{ zhao2024universalsoundseparationselfsupervised }, the authors used an SSL model trained with an MAE loss to encode the audio mixture while using a pre-trained sound event detection (SED) to derive target sound embeddings from enrollment. In our work, we explore using an audio foundation model for TSE systems that can use both class labels and enrollment clues. Besides, we use the same pre-trained foundation model to encode the mixture and the enrollment, removing the need for the SED system and simplifying the system. 
Note that our work also shares similarities with our previous study \cite{ junyi } that used SSL models for target speech extraction. In this paper, we explore whether similar ideas can be effective for the extraction of arbitrary sounds using a mixed target sound class/enrollment TSE, for both online and offline models.


\section{TSE System}
TSE aims to extract a desired sound $\mathbf{s}^{\text{tgt}} \in \mathbb{R}^{1\times T}$ from a sound mixture  $\mathbf{x}\in \mathbb{R}^{1 \times T}$ using a clue $\mathbf{o}$;
\begin{equation}
    \hat{\mathbf{s}}^{\text{tgt}} = \mathrm{TSE}(\mathbf{x}, \mathbf{o};\theta),
\end{equation}
where $\mathrm{TSE}(\cdot)$ is a neural network with parameters $\theta$. The clue $\mathbf{o}$ is either a one-hot vector, $\mathbf{o}^{\text{1-hot}}$ representing the target sound class or an enrollment of the desired sound to extract, $\mathbf{o}^{\text{enroll}}$.
Class label clues, $\mathbf{o}^{\text{1-hot}} \in \mathbb{R}^{C}$, are restricted to the classes in the training set. However, with enrollment clues, we can extend the extraction to sound classes unseen during training \cite{soundbeam}.

\begin{figure}
    \centering
    \includegraphics[width=1.15\linewidth]{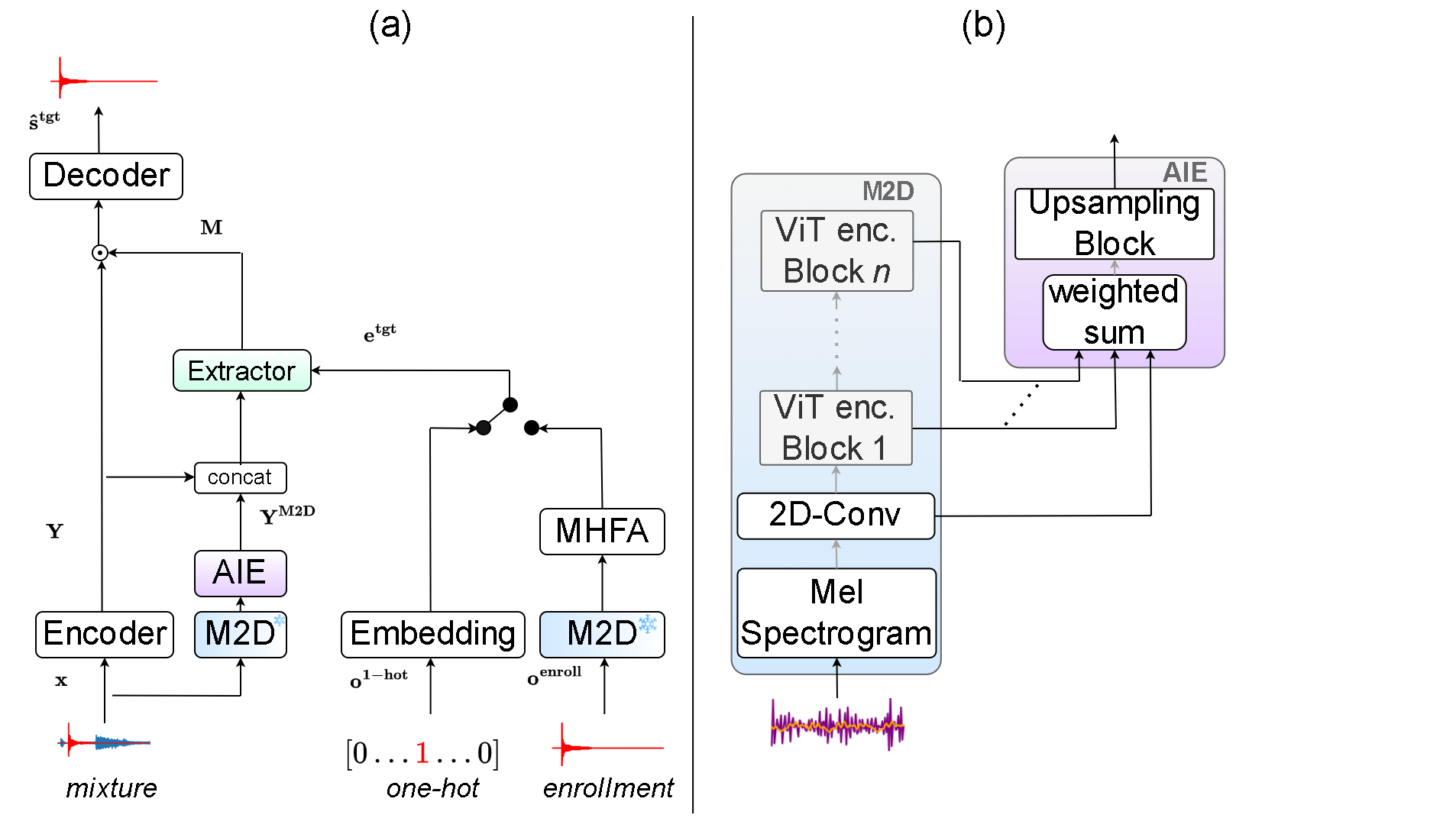}
    \caption{a) Generic TSE system with SSL model and b) M2D model and AIE module.}
    \label{fig:aie}
    \vspace{-4mm}
\end{figure}

\subsection{Model Architecture}
A typical neural TSE system consists of an encoder-decoder model with a mask estimation network or extractor and a clue encoder module, which derives target sound embeddings from the clues. 
The encoder processes the input signal $\mathbf{Y} = \text{Encoder}(\mathbf{x})$, where $\mathbf{Y}\in \mathbb{R}^{T'\times D}$ is the encoded mixture and $T'$ is the time dimension after the encoding, and $D$ the feature dimension. Then, the extractor predicts a mask conditioned by the target sound embedding vector $\mathbf{M} = \text{Extractor}(\mathbf{Y}, \mathbf{e}^{\text{tgt}})$, where $\mathbf{M}\in \mathbb{R}^{T'\times D}$ is the estimated mask and $\mathbf{e}^{\text{tgt}}$ is the target sound embedding. The encoded mixture $\mathbf{Y}$ is multiplied by the mask and converted back into a signal $\mathbf{\hat{s}^{\text{tgt}}}\in \mathbb{R}^{1\times T}$ with the decoder $\mathbf{\hat{s}^{\text{tgt}}} = \text{Decoder}(\mathbf{M} \odot \mathbf{Y})$, to obtain the extracted signal.

We obtain the target sound embedding, $\mathbf{e}^{\text{tgt}}$, by processing the target sound clue with the clue encoder. For class label clues, the clue encoder consists of an embedding layer, and for enrollment clues, it is a neural network combined with an average pooling layer.

\subsection{Training Loss}
All network parameters, including the encoder, decoder, extractor, and clue encoder modules are jointly trained from scratch using the following loss:
\begin{align}
    \mathcal{L}(\theta) = & \alpha \mathcal{L}^{\text{ext}}\left( \mathrm{TSE}(\mathbf{x}, \mathbf{o}^{\text{1-hot}};\theta), \mathbf{s}^{\text{tgt}} \right) \nonumber\\
    &+ (1-\alpha) \mathcal{L}^{\text{ext}}\left(\mathrm{TSE}(\mathbf{x}, \mathbf{o}^{\text{enroll}};\theta), \mathbf{s}^{\text{tgt}} \right),  
    \label{eq:mt_loss}
\end{align}
where $\mathcal{L}^{\text{ext}}$ is the extraction loss, which here is the weighted sum of the signal-to-noise ratio (SNR) and scale-invariant SNR (SI-SNR) losses following \cite{waveformer}. $\alpha$ is a multi-task weight that we fix to $\alpha=0.5$ in all experiments following our prior works \cite{soundbeam}. This loss enables us to jointly train a model for class label and enrollment clues, ensuring good performance thanks to the multi-task training effect and generalization to unseen classes with enrollment clues.

\subsection{Baseline Systems}
In this work, we use two models as our baseline systems, SoundBeam \cite{soundbeam}, and Waveformer \cite{waveformer}, which mainly differ by the implementation of the extractor. 
Both systems use 1D convolution and deconvolution layers for the encoder and decoders. SoundBeam is an offline system based on Conv-TasNet\cite{tasnet}. Waveformer \cite{waveformer} is a lightweight online TSE model that uses dilated convolution and multi-head attention layers for the extractor. 
The original Waveformer used only class label clues. We train it here using both class label and enrollment clues as SoundBeam.
For both SoundBeam and Waveformer models, we use one convolution block of TasNet for the enrollment clue encoder, followed by an average pooling layer.

\section{M2D Model}
\label{sec:m2d}
Let us briefly review the M2D model, before explaining how to use it for TSE in the next Section. 
We use the M2D model pre-trained on AudioSet \cite{audioset} as a pre-trained foundation model. We use the M2D variant, denoted M2D-AS in \cite{m2d_taslp}, where the model is trained by combining an SSL objective with an additional supervised loss of the label prediction of AudioSet audio clips. The SSL objective consists of an improved MAE training objective \cite{m2d_taslp}, which is to predict training references consisting of the features of the masked portion of the input signal using the signal only from the unmasked portion available. Conventional training schemes use the whole input signal for the training reference; thus, the reference could be contaminated with the prediction input. In contrast, M2D's training uses only the masked portion of the input signal for the training reference, preventing contamination that could ease the training objective. 
This encourages the model to learn a more meaningful local representation of the signal, which can be beneficial for TSE. The double training objective of M2D should be particularly suited for TSE, because the TSE problem requires reconstructing the signal (related to the MAE objective) and identifying the source (related to the label prediction loss).

The network architecture of M2D follows that of the vision transformer (ViT)\cite{vit} as shown in Fig. \ref{fig:aie}-(b), which consists of a 2D convolutional (2D-Conv) layer followed by encoder blocks, which we call ViT encoder blocks in the following. The input of the M2D model consists of mel spectrogram features, extracted with a window length of 25 msec and shift of 10 msec The 2D-Conv layer splits the input spectrogram and linearly projects it into tokens. We use M2D with a patch size of $80\times2$ that splits the input every 20 msec, leading to a frame rate of 50 frames per second. 
The model is trained on 2,005,132 audio clips of 10s from AudioSet (5,569 h), cropped into segments of 6 sec.

\section{Proposed: SoundBeam Meets M2D}

Figure \ref{fig:aie} shows a diagram of the proposed TSE system using the pre-trained M2D model.
We use the M2D model to process both the mixture and the enrollment in a similar way as \cite{junyi}.

\subsection{M2D-based Enrollment Clue Encoder}
\label{ssec:m2d_enroll}
We use the pre-trained M2D model to obtain target sound embeddings. We use here an advanced attentive pooling, named multi-head factorized attentive pooling (MHFA), which was first proposed for speaker verification tasks \cite{peng2023attention}. 

MHFA performs attentive pooling to map the enrollment into a fixed-dimension enrollment vector. It generates two distinct weighted-sum of the internal layers of the M2D model. The first one serves as values and the second as keys to generate attention weights for the attentive pooling layer. Finally, the output of the pooling layer is projected to a lower-dimensional space using a linear layer.
MHFA showed promising results for speaker verification \cite{peng2023attention} and target speech extraction \cite{junyi}.

\subsection{M2D for Mixture Encoder}
\label{ssec:m2d_mixture}
To exploit the rich feature representation of the SSL model, we propose to concatenate the encoded mixture, $\mathbf{Y}$, with features obtained from the pre-trained M2D model, $\mathbf{Y}^{\text{M2D}}$. It is well-known that different layers of a pre-trained SSL model capture different types of information. Consequently, we use a weighted sum of the output of all intermediate processing blocks of the M2D model (the first CNN block and each ViT encoder block). In addition, the time resolution of the M2D model and that of the encoder typically used for TSE differ. We thus use a deconvolution layer to upsample the feature representation obtained from the SSL model.
The AIE performs these two operations, weighted-sum and upsampling with deconvolution, as:
\begin{align}
   \mathbf{Y}^{\text{M2D}} = \mathrm{DeConv}\left(\sum_i w_i \mathbf{Z}_i\right),
   \label{eq:aie}
\end{align}
where $\mathbf{Z}_i$ are the output of the $i$-th processing block of the SSL model and $w_i$ are learnable weights constraint so that $\sum_i w_i =1$. 

Note that in \cite{junyi}, we used
a cascade of deconvolution layers to handle the different time-resolutions of the convolutional layers of the speech SSL models \cite{Hsu2021HuBERT,wavlm}.
Here, since the M2D model downsamples the signal using a single convolutional layer, we can upsample all layers with the same deconvolution layers. Another distinction is that the M2D model adds a learnable [CLS] token at the beginning of the sequence to learn global embeddings. We thus pad the output sequence of the CNN layer with a zero to adjust its length with that of the output of the transformer layers.

\subsection{Causal Extension}
\label{ssec:m2d_causal}
Most of our investigations are performed with the offline SoundBeam model. For online processing, we could implement an online version of SoundBeam following the online ConvTasNet configuration \cite{tasnet}. However, we used instead the lightweight Waveformer model, which has shown promising results for online TSE \cite{waveformer,semhear}.

Note that the M2D model is non-causal because of the attention and the layer normalization modules, and thus unsuitable for online processing. For online TSE, we propose modifying the network architecture by replacing attention and layer normalization of the ViT encoder blocks with causal alternatives, i.e., masked attention and cumulative layer normalization \cite{tasnet}. Here, we simply modify an already trained non-causal M2D model to avoid re-training the M2D model. We recognize that this choice may not be optimal, leading to potentially weaker representations. However, we hope that a TSE system can still learn to capture meaningful information from the feature representation obtained from this causal M2D model.

\section{Experimental Settings}

\subsection{Datasets} \label{sec:dataset}
We used the dataset proposed by Semantic Hearing \cite{semhear}, which consists of simulated reverberant binaural mixtures of three to four sound events added to urban background noise.
This dataset leverages 20 sound classes from FSD50K \cite{FonsecaFPFS22} (general-purpose), ESC-50 \cite{Piczak15} (environmental sounds), MUSDB18 \cite{musdb18} and noise files for the DISCO dataset \cite{disco}. 
The background sounds were taken from TAU Urban Acoustic Scenes 2019 \cite{MesarosHV18}. Binaural mixtures were generated using sound source signals with room impulse responses (RIRs) and head-related transfer functions (HRTFs). We used HRTFS from 43 subjects from the CIPIC corpus\cite{algazi2001cipic}, and real and simulated RIRs from three corpora\cite{SBSBRIR,IoSR_Surrey_2016,IoSR_Surrey_2023}. The sampling frequency of all signals was 16 kHz.
To train single-channel models in this work, we average the right and left channels of the mixtures and enrollments.

For training, the data is mixed on the fly with Scaper \cite{salamon2017scaper}. From each mixture, we extracted two foreground sources. 
The training was done with 6-sec mixtures. The number of training, validation, and testing mixtures were 100K, 1K, and 10K, respectively.
During training, the enrollment cues are chosen by randomly sampling audio files from the target sound class, different from the ones used in each mixture. For validation and test, the enrollments are fixed for each mixture. Note that the enrollments are also reverberant.

\subsection{System Configuration}
We experiment with the following four models:

\subsubsection{Baseline SoundBeam (Offline Model)}
We used one 1D-Conv and 1D-DeConv layers for the encoder and decoder, respectively, with a kernel size of 16 and a stride of 8. For the extractor, we used 512 filters, 8 blocks, 3 repeats, and a global layer norm. 
For the class label clue embedding network, we used 2 fully connected layers with layer normalization following \cite{semhear}. For the enrollment embedding network, we used a kernel size of 40, a stride of 20, and an embedding size $D=256$.

\subsubsection{SoundBeam with M2D}
We used the same configuration as the baseline SoundBeam model but replaced the enrollment clue encoder with the M2D-based one as explained in Subsection \ref{ssec:m2d_enroll}.
In addition, we also experimented using the M2D model to process the mixtures. In that case, we used the AIE module described in Subsection \ref{ssec:m2d_mixture}, where we used two 1D-DeConv layers of kernels $\{ 2, 25 \}$ and strides of $\{ 2, 20 \}$ to implement the deconvolution operation of Eq. \eqref{eq:aie}.

\subsubsection{Baseline Waveformer (Online Model)}
We used the same setting as \cite{semhear}. The encoder and decoder consist 1D-Conv and 1D-DeConv layers with stride of $L=32$ samples and a kernel size of $K=3L$; therefore, the lookahead of the model is 32 frames, and the downsampling factor is 8. For the extractor, we used 10 DCC layers with a kernel size of $K$=3 and dilation factors set to $\{ 2^0, 2^1, ..., 2^9 \}$. The masker decoder consisted of one MHA layer with 8 heads. 
We used the same clue networks and parameters as for SoundBeam.

\subsubsection{Waveformer with M2D}
We used the same configuration as the baseline Wavefomer model but replaced the enrollment clue encoder with the M2D-based one as explained in Subsection \ref{ssec:m2d_enroll}. Processig the enrollment can be performed offline, and we thus used the original non-causal M2D model to process the enrollment.

In contrast, to keep the model online, we used the causal implementation of M2D, explained in Section \ref{ssec:m2d_causal}, to process the mixture. Here, we implemented the deconvolutional layer of Eq. \eqref{eq:aie} with four 1D-DeConv layers of kernels $\{ 2, 2, 2, 1 \}$ and strides of $\{ 2, 2, 2, 5 \}$. 
Then, we added a bottleneck CNN layer of kernel and stride 1 that processes the encoded mixture concatenated with the output of the AIE CNN block.

\subsubsection{Other Settings}
We trained all models for 80 epochs with four GPUs using a batch size of 24. We chose the models that achieved the best SI-SNR score on the cross-validation set. 

During inference, we perform extraction using either class label or enrollment clues. For each mixture, we extract two target sounds for the evaluation.
We report results in terms of SNR and failure rate (FR), which is the percentage of samples with SNR improvement (SNRi) below 1~dB. 

\subsection{Experiments with SoundBeam (Offline TSE)}

\begin{table}[t]
\caption{TSE performance in terms of SNRi for offline SoundBeam model with and without M2D. The SNR of the mixture is -0.4 dB. \cmark indicates that the M2D model is used for the enrollment or the mixture.}
    \centering
    \scriptsize
    \begin{tabular}{
    ccccc
    }
    \toprule
 \multicolumn{2}{c}{Use M2D model for} & Clue used for &  SNRi $\uparrow$  & FR $\downarrow$ \\
          enrollment & mixture & inference &   [dB] &  [\%]\\
         \midrule


         
          \multirow{2}{*}{-} & \multirow{2}{*}{-} & class label & 9.48 & 0.05 \\
          & & enrollment & 7.72 &  0.19 \\
 
%


\midrule
           \multirow{2}{*}{\cmark} & \multirow{2}{*}{-} & class label & 9.52 & 0.05 \\
         & & enrollment & 9.13 & 0.09 \\
\midrule
         
        \multirow{2}{*}{\cmark} & \multirow{2}{*}{\cmark} & class label & 10.49 & 0.03 \\
         &  & enrollment & 9.86 & 0.08 \\

        \bottomrule
    \end{tabular}
    \label{tab:results_soundbeam}
    \vspace{-4mm}
\end{table}

Table \ref{tab:results_soundbeam} compares the baseline SoundBeam model with two variants using M2D to process the enrollment and the mixture when using class label or enrollment clues for inference. We observe that using M2D model greatly improves the extraction performance when using enrollment clues for inference (by more than 2 dB). This can be attributed in part to better target sound identification, as shown by the reduction in terms of FR. Note that the best performance was achieved when exploiting M2D for both the enrollment clues encoder and for processing the mixture.

\begin{figure}[tb]
    \centering
    \begin{minipage}{.5\textwidth}
        \centering
        \includegraphics[width=0.75\linewidth]{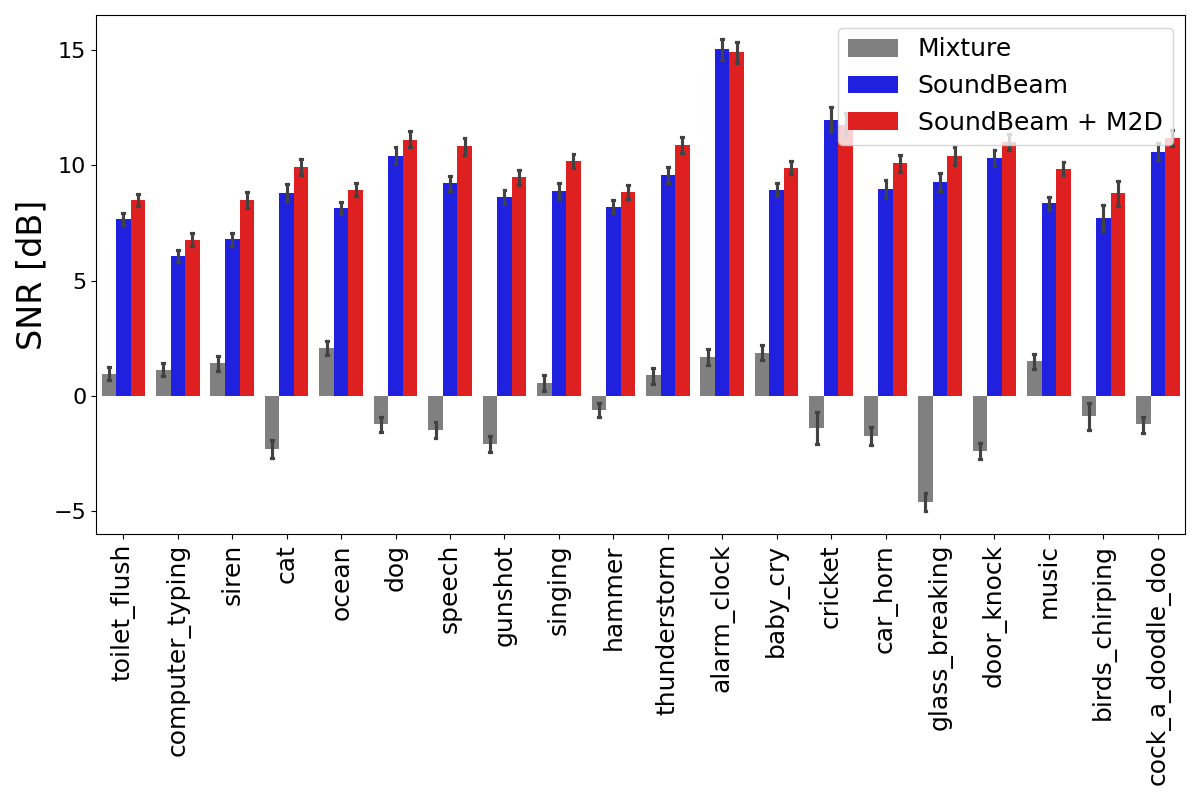}
    \end{minipage}
    \\
    \begin{minipage}{0.5\textwidth}
        \centering
        \includegraphics[width=0.75\linewidth]{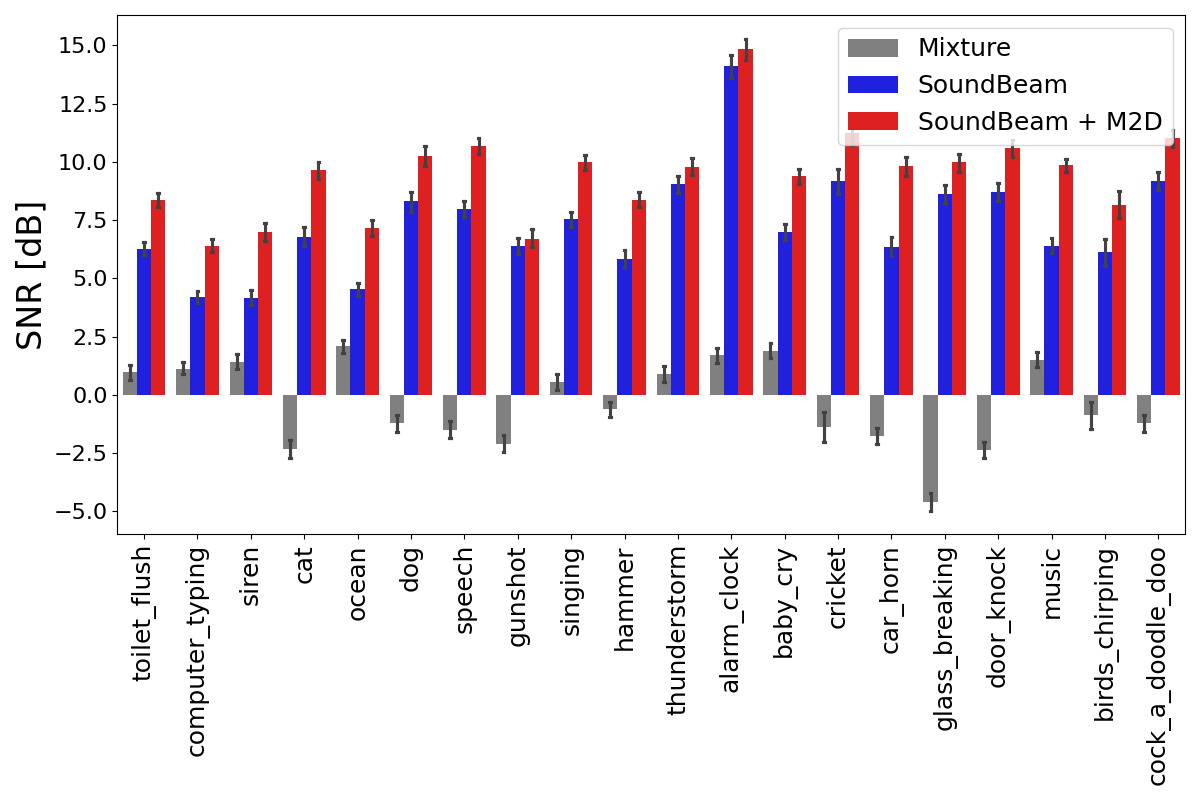}
    \end{minipage}
    \vspace{-2mm}
    \caption{SNR for different target sound classes using class label (top figure) or enrollment (bottom figure) clues.}
    \label{fig:res_per_class}
\end{figure}

When using the M2D model to process the mixture, the performance of class label clue-based inference also improves by 1 dB. These results confirm that using the M2D model is promising for the TSE task. 
Figure \ref{fig:res_per_class} shows the performance for each of the 20 target sound classes. It confirms that using M2D provides consistent improvements for almost all target sound classes.

\subsection{Experiments with Waveformer (Online TSE)}

\begin{table}[t]
\caption{TSE performance in terms of SNRi for online Waveformer model with and without M2D.}
    \centering
    \scriptsize
    \begin{tabular}{
    ccccc
    }
    \toprule
 \multicolumn{2}{c}{Use M2D model for} & Clue used for &  SNRi $\uparrow$  & FR $\downarrow$ \\
          enrollment & mixture & inference &   [dB] &  [\%]\\
         \midrule



         \multirow{2}{*}{-} & \multirow{2}{*}{-} & class label & 7.75 & 0.07\\
          &  & enrollment & 6.01 & 0.23\\

         \midrule

         \multirow{2}{*}{\cmark} & \multirow{2}{*}{-} & class label & 7.43 & 0.09 \\
           & & enrollment & 7.01 & 0.12 \\

          \midrule

          \multirow{2}{*}{\cmark} & \multirow{2}{*}{\cmark} & class label & 7.73 & 0.07 \\
          & & enrollment & 7.17 & 0.10 \\

        \bottomrule
    \end{tabular}
    \label{tab:results_waveformer}
    \vspace{-4mm}
\end{table}

Table \ref{tab:results_waveformer} shows extraction performance for experiments based on the Waveformer online model. As for SoundBeam experiments, we also observe that using M2D leads to improved performance when using enrollment clues for inference, but not when using class label clues. Nevertheless, the resultant model becomes more practical as it can perform both class label and enrollment clue-based TSE, with a similar level of performance.

Note that the lack of improved performance when using class label clues may be caused by the weak representation of the mixture provided by our causal version of the M2D model.
This issue may be alleviated by pre-training a causal M2D model directly or retraining its parameters jointly with TSE. These investigations will be part of our future work.

\section{Conclusions}
We have proposed a new TSE system that combines the SoundBeam model with the strong M2D audio foundation model. We showed that the M2D model can significantly improve performance using class label and enrollment clues, by providing more discriminative target sound embeddings and better representation of the mixture.
In future work, we would like to explore causal implementation or retraining of M2D as well as lightweight configuration to boost the performance of online TSE.
\bibliographystyle{IEEEbib}
\bibliography{refs}

\end{document}